\documentclass[fleqn,usenatbib]{mnras}

\usepackage{newtxtext,newtxmath}
\usepackage[T1]{fontenc}

\DeclareRobustCommand{\VAN}[3]{#2}
\let\VANthebibliography\thebibliography
\def\thebibliography{\DeclareRobustCommand{\VAN}[3]{##3}\VANthebibliography}

\usepackage{graphicx}	% Including figure files
\usepackage{amsmath}	% Advanced maths commands
\usepackage{xcolor}

\newcommand{\NE}{\bar{N}\bar{\xi}}

\title[VPF in Voids]{Void Probability Function inside cosmic voids: evidence for hierarchical scaling of high-order correlations in real space}

\author[Dávila-Kurbán et al.]{
Federico Dávila-Kurbán,$^{1}$\thanks{E-mail: fdavilakurban@unc.edu.ar (FDK)}
Andr\'es N. Ruiz,$^{1,2}$
Dante J. Paz$^{1,2}$
and Diego Garcia Lambas$^{1,2}$
\\
$^{1}$Instituto de Astronomía Te\'orica y Experimental, CONICET-UNC, Laprida 854, X5000BGR, C\'ordoba, Argentina\\
$^{2}$Observatorio Astron\'onmico, Universidad Nacional de C\'ordoba, Laprida 854, X5000BGR, C\'ordoba, Argentina\\
}

\date{Accepted XXX. Received YYY; in original form ZZZ}

\pubyear{2023}

\begin{document}
\label{firstpage}
\pagerange{\pageref{firstpage}--\pageref{lastpage}}
\maketitle

\begin{abstract}
We compare the reduced void probability function (VPF) inside and outside of cosmic voids in the TNG300-1 simulation, both in real and simulated redshift space. 
The VPF is a special case of the counts-in-cells approach for extracting information of high-order clustering that is crucial for a full understanding of the distribution of galaxies. Previous studies have validated the hierarchical scaling paradigm of galaxy clustering moments, in good agreement with the ``negative binomial'' model, in redshift surveys, but have also reported that this paradigm is not valid in real space.
However, in this work we find that hierarchical scaling can indeed be found in real space inside cosmic voids. 
This is well fitted by the negative binomial model.
We find this result to be robust against changes in void identification, galaxy mass, random dilutions, and redshift. 
We also obtain that the VPF in real space at high redshift approaches the negative binomial model, and therefore it is similar to the VPF inside voids at the present time.
This study points, for the first time, towards evidence of hierarchical scaling of high-order clustering of galaxies in real space inside voids,
preserving the pristine structure formation processes of the Universe.
\end{abstract}


%%%%%%%%%%%%%%%%%%%%%%%%%%%%%%%%%%%%%%%%%%%%%%%%%%

%%%%%%%%%%%%%%%%% BODY OF PAPER %%%%%%%%%%%%%%%%%%

\section{Introduction}
\label{sec:intro}

The Void Probability Function (VPF) estimates the probability of finding zero galaxies in a given volume, and it is a special case of the Counts-in-Cells (CiC) analysis. This approach to clustering statistics has been used to estimate higher-order correlation functions \citep{ColesJones1991,Baugh2004a,Croton2004}, which are increasingly difficult to obtain, in order to extract more information than is available in the more traditionally used two-point correlation function. A further understanding of these higher-orders of correlation is needed for a full description of the clustering of galaxies and the small- and large-scale structures it originates.
This work aims to characterise the VPF inside and outside cosmic voids, in real and redshift space, and to study what information this underused statistic can provide.

An alternative method for characterizing the clustering of a sample has been the CiC analysis, which yields the count probability distribution function (CPDF). This approach focusses on the computation of the probability of finding $N$ objects in a volume $V$, $P_N(V)$. If one can calculate this statistic for all $N$, then in principle one can fully characterize the distribution of the sample. The VPF is a particular case of this approach where $N=0$, and its main appeal has been its theoretical relation to all high-order correlation functions \citep{White1979,Sharp1981,Fry1986,BalianSchaeffer1989,Gaztanaga1993,Vogeley1994,Benson2003a}. Whenever this relation is valid, the VPF can provide a pathway to the information encoded in every correlation order that would otherwise be very costly to obtain.

The calculation of the VPF of a given set of data has been often contrasted in the literature with known hierarchical clustering models \citep[e.g.][]{Croton2004models,Mekjian2007,Fry2013}. 
These models can be understood as recepies for writing the scaling coefficients that parameterize the hierarchy of the correlation functions.
In particular the Negative Binomial model (NB, hereafter) has been shown to be a very good fit for the VPF of galaxies in redshift surveys \citep[][and references above]{ElizaldeGaztanaga1992,Conroy2005,Croton2006a}, while the Thermodynamic model provides a less accurate fit for some dark matter samples and galaxies in certain cosmological models \citep{Vogeley1994,Croton2004models}.
However, despite evidence of the success of the NB model as a fit for the VPF of galaxies in redshift surveys, and therefore confirming the hierarchical scaling of the clustering correlations, one main problem remains: this behaviour does not hold in real space. 
For example, \cite{Vogeley1994} simulate real and redshift space with different cosmologies and confirms the idea (previously reported as well by \citealt{Lahav1993}) that the hierarchical scaling is only valid in redshift space, i.e. when the clustering pattern is distorted by peculiar velocities. 
However, \cite{Croton2006a} find that fainter and redder galaxies move away from the NB model and so disputes the idea that the distortion of peculiar velocities is the main reason for the success of the NB model.

Similar reports for the success of the NB model have been presented for the entire CiC treatment. Using data from the Sloan Digital Sky Survey (SDSS), \cite{Hurtado-Gil2017} conclude that the NB model is the best fit for the CiC distribution: not only for the probability $P_0$ of finding zero galaxies but also for the probability $P_N$ of finding $N$ galaxies with $N\geq 1$, while the lognormal distribution (a model with similar behaviour to that of the Thermodynamic) is, with some modifications, a justifiable alternative for cells spanning large scales.

However, there are some deviations from the NB model. \cite{Croton2006a} find that red galaxies of the SDSS have stronger clustering strength than blue galaxies at the orders they measured and, therefore, stray upwards from the NB fit. 
In agreement with this result
\cite{Tinker2008} discovered that occupation functions resulting in correlation functions that significantly deviate from a power law tend to stray from the NB model. They argue that the reason for this departure is the prevalence of satellite galaxies within the red occupation function, which leads to a distinct shift from the one- to two-halo regime. On the other hand, the correlation function for the blue sample closely follows a power law, and is therefore well fitted by the NB model.%

It is also important to note that the Ansatz of hierarchical scaling is obtained with two key assumptions: self-similarity and stable clustering \citep{Bernardeau2002}. Stable clustering implies that the coefficients are constant with time, while self-similarity leads to coefficients that are constant with scale. 
Both of these approximations are simplistic, but they have been useful in understanding the behavior of small-scale correlations.
Therefore, it is to be expected that the Ansatz of hierarchical scaling is true only in regimes in which these approximations have reasonable validity.

Authors such as \cite{Conroy2005} and \cite{Tinker2006a} argue that the VPF is completely determined by the number density and the volume-averaged two-point correlation function of the sample, so it does not necessarily provide more information than these statistics and should be used as a complementary method. 
Nevertheless, the CiC is a simple but powerful statistical approach to characterise the distribution of galaxies in space, since it contains statistical information about voids and subdense regions, clusters of various sizes and shapes, filaments, about the probability of finding an arbitrary number of neighbours around random positions, about counts of galaxies in randomly positioned cells with abritrary shapes and random sizes, and about the correlation function of galaxies of all orders \citep[][and references therein]{YangSaslaw2011}. 
And despite the wealth of information about galaxy clustering this statistic contains, it has not received as much attention as more traditional clustering statistics such as the two-point correlation function, not in small part because of the difficulty in extracting from it a straightforward interpretation.
The VPF alone may be insufficient to fully describe the galaxy distribution, but it is a complementary statistic that is part of the CiC approach which holds great potential. 
Moreover, the void distribution is relatively insensitive to information on large cell sizes because large cells are unlikely to be completely empty \citep{YangSaslaw2011}. This makes cosmic voids a particularly good environment for using the VPF.

With this work, we aim to shed some light on the information encoded in the void probability function on different environments, in particular inside void regions in comparison with the general Universe. We focus on how this statistic is sensitive to different tracers, different redshifts and the different expansion history provided by void environments.

Following this introduction, Sec.~\ref{sec:methods} provides a brief overview of the statistics needed for the calculations performed, including necessary definitions and uncertainties estimations.
Sec.~\ref{sec:data} details the simulation used and the void identification algorithm. Results are shown and analysed in Sec.~\ref{sec:results} and finally, in Sec.~\ref{sec:conclusions}, we summarize this work while discussing our findings.

\section{Methods}
\label{sec:methods}

\subsection{Statistics Overview}
Given a distribution of galaxies, the count probability distribution function (CPDF), $P_N(R)$, describes the probability of finding $N$ galaxies in a radius $R$. It can be shown that the VPF, or the probability of finding zero galaxies $P_0(R)$, can be related to the hierarchy of p-order correlation functions \citep{White1979}:

\begin{equation}
    P_0(R)=\mathrm{exp}\bigg\{ 
    \sum\limits_{p=1}^\infty \frac{[-\Bar{N}(R)]^p}{p!}\Bar{\xi}_p(R)
    \bigg\},
    \label{eq:vpf_original}
\end{equation}
where $\Bar{N}(R)$ is the mean number of galaxies in a sphere with radius $R$, and $\Bar{\xi}_p$ is the volume averaged $p$-th order correlation function:

\begin{equation}
    \Bar{\xi}_p = \frac{\int\xi_pdV}{\int dV}.
\end{equation}

The expression in Eq.~\ref{eq:vpf_original} can be simplified when inserting the ``hierarchical scaling'' Ansatz:

\begin{equation}
    \bar{\xi}_p(R) = S_p\bar{\xi_2}^{p-1}(R),
    \label{eq:hierarchicalscaling}
\end{equation}

\noindent which then gives the expression:

\begin{equation}
    P_0(R)=\mathrm{exp}\left[ 
    \sum\limits_{p=1}^\infty \frac{[-\Bar{N}(R)]^p}{p!}S_p\bar{\xi_2}^{p-1}(R)
    \right].
    \label{eq:vpf_jerarq}
\end{equation}

The aforementioned Ansatz postulates that the higher-order correlations $\xi_p$ are related to the lowest order of correlation $\xi_2$ (omitting the subscript hereafter for simplicity) by way of constant coefficients $S_p$.
The concept of hierarchical emergence of the higher-order clustering from the two-point correlation function is natural in both perturbation theory and the highly non-linear regime of gravitational clustering \citep{Peebles1980}. This idea has been strongly reinforced by a wealth of observational evidence (see, e.g., the review of \citealt{Bernardeau2002} and references therein).

Finally, it can be seen from Eq.~\ref{eq:vpf_jerarq} that $P_0$ can be written in terms of $\bar N$ and $\bar \xi$ alone. This is the so-called ``scaling variable'', $\NE$. 
This quantity can be roughly thought of as the number of objects in excess with respect to a randomly distributed sample for a given volume. Naturally, this variable $\NE$ has a finite range, depending on the sample; e.g., this number can never grow too large on a sample with no clustering.
\cite{Fry1986} formalizes the concept of writing the VPF in terms of the scaling variable by introducing $\chi$, the \textit{reduced} VPF (RVPF), defined as:

\begin{equation}
    \chi = -\mathrm{ln}(P_0)/\bar{N},
    \label{eq:rvpf_p0}
\end{equation}

\noindent so that when combined with Eq.~\ref{eq:vpf_jerarq} we obtain:

\begin{equation}
    \chi(\bar{N}\bar{\xi})= 
    \sum\limits_{p=1}^\infty \frac{S_p}{p!}(-\bar{N}\bar{\xi})^{p-1}.
    \label{eq:rvpf_NE}
\end{equation}

This expression shows that, under the assumption of the scaling relation, we can anticipate a convergence of different galaxy samples with varying densities and clustering strengths onto a universal curve. This convergence occurs because of the shared dependence on a common scaling variable. If this holds, it also follows that differences in the shape of VPFs of galaxy samples must arise only from differences between their higher-order correlation functions, i.e. differences in their $S_p$.
Conversely, if changing the density, correlation strength or some other parameter of a sample, yields vastly different VPFs, that could indicate that the coefficients $S_p$ are not constant and indeed depend strongly on the parameter that is being tested. This scenario would invalidate the hierarchical Ansatz of Eq.~\ref{eq:hierarchicalscaling}.

Different ways of fixing the scaling coefficients $S_p$ constitute different models of hierarchical scaling. For this work, we will use the two models that more closely fit our data: the negative binomial model, and the thermodynamic model.
Previous works have shown that the NB model is a very good fit for the VPF of galaxies in redshift surveys, while the thermodynamic model has been proposed as a loose fit for some dark matter samples and galaxies in a few obsolete cosmological models \citep{Vogeley1994,Croton2004models,Conroy2005,Croton2006a,Fry2013}.
The RVPFs and scaling coefficients of the NB and thermodynamical models are, respectively:

$$\chi_\mathrm{NB} = \mathrm{ln}(1+\bar{N}\bar{\xi})/\bar{N}\bar{\xi};\; \text{with}\; S_p = (p-1)!,$$ 

and 

$$\chi_\mathrm{T} = [(1 + 2 \bar{N}\bar{\xi})^{1/2} - 1]/\bar{N}\bar{\xi};\; \text{with}\; S_p = (2p - 3)!!.$$

\subsection{Main quantities}
Computing the VPF requires the measurement of three quantities: $\Bar{N}(R)$, $\Bar{\xi}(R)$, and $P_0(R)$. 
These quantities are determined directly from the CiC approach (their dependency on $R$ is hereafter made implicit). The procedure can be summarized as follows: a large number $N_\mathrm{sph}$ of probing spheres of radius $R$ are randomly placed throughout the data, the TNG300-1 in our case, and the number of galaxies in each sphere is counted. Finally, we calculate the probability of finding an empty sphere. This is then repeated for a number of radii in a given range.

Here, we briefly define the main quantities mentioned above. Let $\Bar{N}$ be the mean number of galaxies in a sphere (Eq.~\ref{eq:nbar}), and $P_0$ be the probability that this volume is empty, as shown in Eq.~\ref{eq:p0}, where $N_0$ is the number of spheres with zero galaxies and $N_\mathrm{sph}$, as mentioned above, is the total number of probing spheres.
Finally, let $\Bar{\xi}$ be the variance in the number of galaxies per sphere (Eq.~\ref{eq:xi}):

\begin{equation}
    \Bar{N} = \frac{1}{N_{\mathrm{sph}}}\sum_{i=1}^{N_{\mathrm{sph}}}N_i
    \label{eq:nbar}
\end{equation}

\begin{equation}
    P_0=\frac{N_0}{N_\mathrm{sph}},
    \label{eq:p0}
\end{equation}

\begin{equation}
    \Bar{\xi} = \frac{\overline{(N-\Bar{N})^2}-\Bar{N}}{\Bar{N}^2}.
    \label{eq:xi}
\end{equation}
In the limit of $N_\mathrm{sph}\xrightarrow{}\infty$ the CiC approach to determine the volume-averaged correlation function, $\Bar{\xi}$, is mathematically equivalent to using the Landy-Szalay estimator \citep{Szapudi1998, Conroy2005}.

We studied the sensitivity of these quantities to the total number of spheres used, and find that the quantities converge well for $N_\mathrm{sph}>10^4$ (see Appendix~\ref{sec:appA}). To ensure a reliable estimate of them we use $N_\mathrm{sph}=10^5$ for the calculation throughout the simulation.
The maximum probing radius throughout the box was chosen so that the resulting $\chi(\NE)$ value is close to zero; we found this value to be $r_\mathrm{max}=4h^{-1}$Mpc. This way we probe the largest $\NE$ range possible.

For the calculation of the VPF inside voids, the choice of $N_\mathrm{sph}$ is non-trivial and is dependent on the maximum probing radius if we want to minimize oversampling.
The maximum amount of non-overlapping spherical volumes $V(r)$ that can be fit inside a bigger spherical volume $V(R)$ is $n_\mathrm{max}=V(R)/V(r)=(R/r)^3$ in the limit where $R/r\to\infty$. However, for $1\ll R/r\ll\infty$, the most efficient sphere packing with randomly placed spheres is approximately 64 per cent \citep{JaegerNagel1992}, so we can approximate \mbox{$n_\mathrm{max}\simeq0.64(R/r)^3$}. Considering this relation we can derive our maximum probing radius within voids:
$r$ cannot grow so large so as to make $n$ drop below the needed $10^4$ (see the above Sec.~\ref{sec:uncertainties} and App.~\ref{sec:appA}) across all voids.
We find that $N_\mathrm{sph}=2\times10^4$ is a good compromise between a reliable estimation of the CiC statistics, and a large enough maximum probing radius that minimizes chances of oversampling.

When choosing the smallest void radius of the sample as $R$, this relation yields $r_\mathrm{max}\simeq1.6h^{-1}$Mpc as the largest probing radius within voids. Independent runs (with different random seeds) of the VPF calculation algorithm within voids with this choice of $N_\mathrm{sph}$ (and therefore, of $r_\mathrm{max}$) yields statistically identical results.

Finally, since the simulation is periodic in volume, its center and edges are equally sampled, regardless of the radius of the sampling spheres. In practice we ensure this by extending the simulation periodically to the bounds $[L_\mathrm{box}-R_\mathrm{max},\,L_\mathrm{box}+R_\mathrm{max}]$, where $R_\mathrm{max}$ is the radius of the largest sampling sphere. 
Finally, to simulate redshift space in TNG300-1 we use the distant observer approximation using the $H_0=67.74$ km s$^{-1}$ Mpc$^{-1}$ parameter of the simulation and making the conversion $x^s=x+v_\mathrm{x}/H_0$, where $x$ and $v_\mathrm{x}$ are the coordinate and velocity in a given axis respectively, and apply periodic conditions when appropriate.
We have calculated the statistics presented in this section modifying the three axes of the box and have found that the results are indistinguishable from each other.

\subsection{Uncertainties}
\label{sec:uncertainties}

The uncertainties of $\bar{N}$ and $\bar{\xi}$ were estimated by Jackknife resampling of the total volume into thirds, resulting in 27 equal subvolumes. Subsequently, we used analytically derived expressions to calculate the uncertainty of $P_0$ and $\chi$ \citep{Hamilton1985, Maurogordato1987VoidSegregation, Colombi1994AModel}:

\begin{equation}
    \Delta P_0=\sqrt{\frac{P_0(1-P_0)}{N_\mathrm{sph}}},
\end{equation}

\noindent and

\begin{equation}
    \left(\frac{\Delta \chi}{\chi}\right) \simeq \left| \frac{\Delta P_0}{P_0|\mathrm{ln}P_0|} - \frac{\Delta\bar{N}}{\bar{N}} \right|,
\end{equation}

\noindent since when calculating the uncertainty of $\chi=-\mathrm{ln}(P_\mathrm{0})/\bar{N}$, the denominator and numerator are not independent but almost exactly anti-correlated \citep[see][]{Colombi1994AModel, Fry2013}.

To calculate the statistics within the cosmic voids identified in TNG300-1, we uniformly sampled their spherical volume. 
Jackknife estimates of the statistics were done by ignoring a subset of voids from the complete sample so that we have the same number of jackknife resamplings as in the computation of the VPF throughout the box.

\section{Data}
\label{sec:data}

In this work we employ galaxy data from the \textsc{IllustrisTNG}\footnote{https://www.tng-project.org/} project \citep{Pillepich2018b, Marinacci2018, Naiman2018, Springel2018, Nelson2018, Nelson2019b, Pillepich2019}. \textsc{IllustrisTNG} is a set of cosmological magneto-hydrodynamical simulations obtained with the \textsc{arepo} moving-mesh code \citep{Springel2010}, and adopting a Planck cosmology \citep{Collaboration2016}: $\Omega_\mathrm{m}= 0. 3089$, $\Omega_\mathrm{b}= 0.0486$, $\Omega_\mathrm{\Lambda}= 0.6911$, $\sigma_8= 0.8159$, $n_s= 0.9667$, and $h= 0.6774$. These simulations present comprehensive models for the physics of galaxy formation, and improve upon its predecessor, \textsc{Illustris}, by including magnetic fields and improving models of the galactic wind and AGN feedback. The \textsc{IllustrisTNG} project encompasses three different volumes with identical initial conditions and physical models: TNG50, TNG100 and TNG300. 
In particular, we employ TNG300-1, which has periodic box of 205$h^{-1}$Mpc, the largest box with the highest resolution in the suite. 
The halos (clusters) and subhalos (galaxies) are found with a standard ``friends of friends'' (FoF) algorithm with link length $b=0.2$ (in units of the mean interparticle spacing) run on the dark matter particles, and with the \textsc{subfind} algorithm \citep{Springel2001}, respectively. The latter detects substructure within clusters and defines locally overdense, self-linked particle clusters, where the baryonic component in the substructure is defined as a galaxy.
We analyze this simulation at redshift $z=0, 0.5, 1$ and 2, and use galaxies in the mass range $10^9\leq M/M_\odot\leq10^{13}$.

The identification of voids in the simulation follows the algorithm described in \cite{Ruiz2015}, a modified version of previous algorithms presented in \cite{Padilla2005} and \cite{Ceccarelli2006}. 
The algorithm estimates the density profile with a Voronoi tessellation over density tracers, which in this work are galaxies from the TNG300-1 simulation. Subdense regions are obtained by selecting Voronoi cells below a density threshold and selected as cosmic void candidates. Centered on these cells, the integrated density contrast $\Delta(r)$ is calculated at increasing values of $r$. Void candidates are selected as the largest spheres satisfying the condition $\Delta(R_\mathrm{v}) = -0.9$ where $R_\mathrm{v}$ is the radius of the void. Next, the centers of the voids are randomly shifted so that the spheres can grow. This is done because the algorithm is likely to produce spherical voids where their shells do not accurately match the surrounding structures, and the recentering procedure provides structures with edges that better match the surrounding local density field. Finally, the catalog of voids comprises the largest subdense, non-overlapping spheres of radius $R_\mathrm{v}$. 

The tracers we use to identify voids are galaxies of TNG300-1 with a lower total mass cutoff of $M\geq10^{11}M_\odot$ to emulate observable voids. For the proper VPF calculations, as stated above, we include galaxies down to a $M\geq10^9M_\odot$ threshold.
After applying this algorithm to TNG300-1 and eliminating shot-noise voids, we are left with a sample of 174 voids in real space and 182 in redshift space with radii in the range of 9-18 $h^{-1}$Mpc.

\section{Results}
\label{sec:results}

\begin{figure*}
    \centering
    \includegraphics[width=\textwidth]{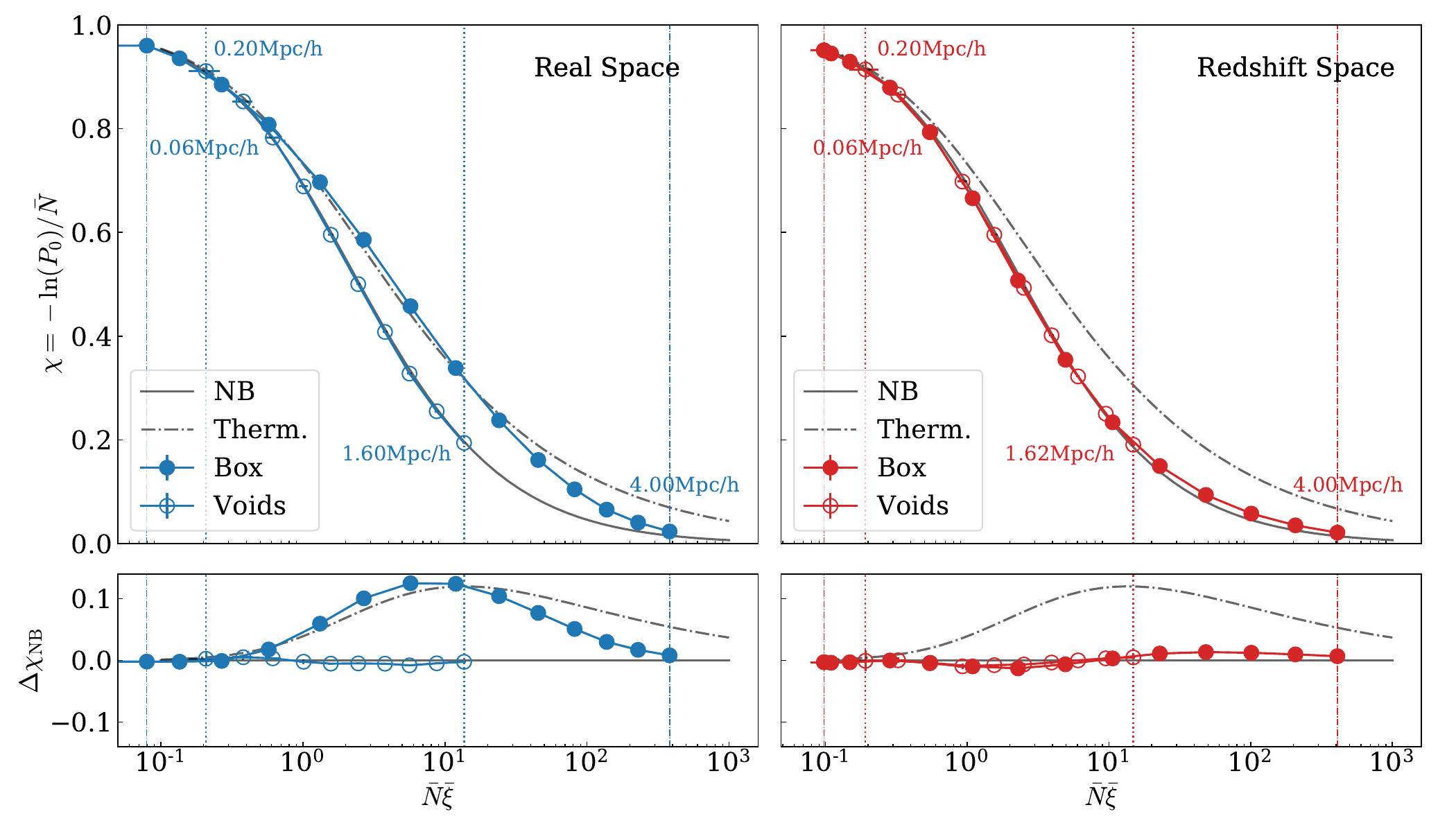}
    \caption{Reduced VPF, $\chi(\Bar{N}\Bar{\xi})$, inside (empty circles) and outside (filled circles) of cosmic voids, in real and redshift space (blue and red symbols on left and right panels, respectively). 
    Vertical lines with scales corresponding to the first and last data values of $\NE$ for each sample are provided for reference. 
    The negative binomial model of clustering (NB, grey solid line) is known to be a good fit for the VPF of galaxies in redshift space. However, we find a striking agreement between this model and the VPF of galaxies within voids in real space (empty blue circles), which is an indication of a previously unknown clustering scaling relation in real space.
    The bottom panels show the difference between each of the reduced VPFs with that of the NB model, where the agreement between the data and the fit can be seen more clearly.}
    \label{fig:chi_vs_NXi_bothspaces_invoid+box}
\end{figure*}

\begin{figure*}
    \centering
    \includegraphics[width=\textwidth]{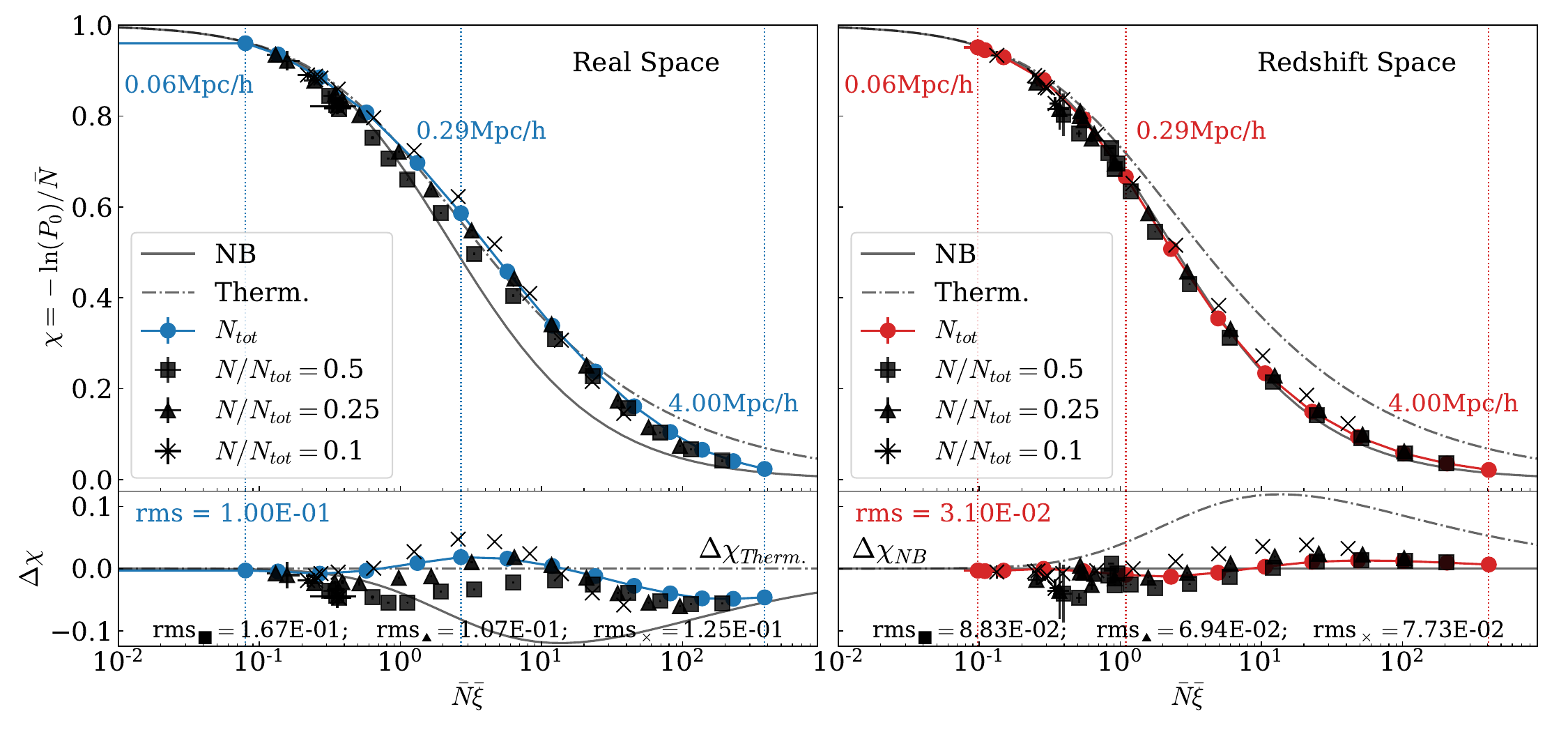}
    \includegraphics[width=\textwidth]{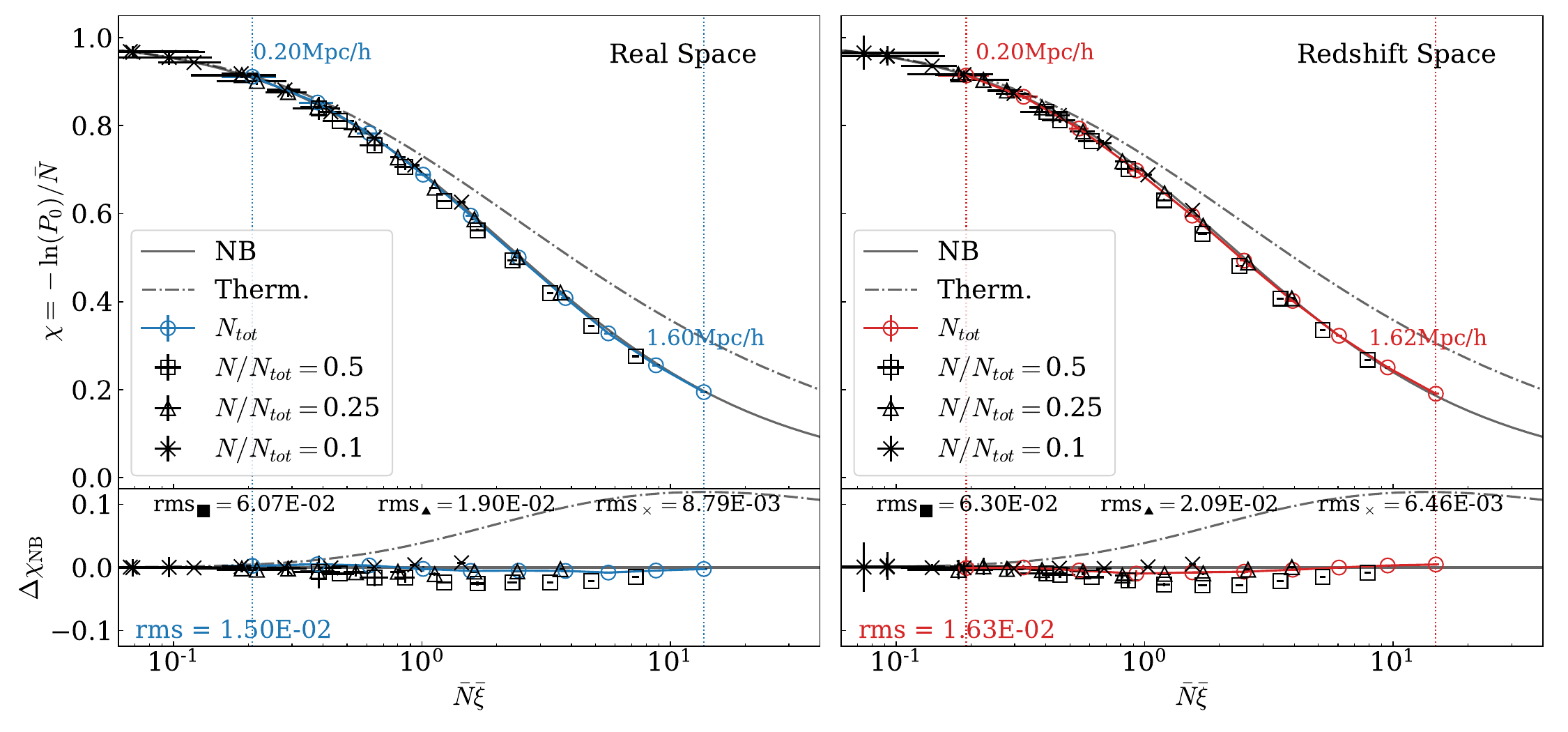}
    \caption{Reduced VPF of galaxies within cosmic voids in TNG300-1 in redshift (right panels) and real space (left panels) with random dilutions of 50, 25, and 10 per cent. The bottom panels show the subtractions between the RVPF of the samples with the RVPF of the thermodynamic model ($\Delta\chi_\mathrm{Therm.}$, bottom left panel) and with that of the negative binomial model ($\Delta\chi_\mathrm{NB}$, bottom right panel). Contrary to previous results, the low scatter in the diluted RVPFs in the lower left panel indicates a possible detection of hierarchical scaling of galaxies within cosmic voids in real space.}
    \label{fig:chi_vs_NXi_dilute_bothspaces_invoid+box}
\end{figure*}

As a general reference, we show results in real space in blue colours and redshift space in red colours. Statistics corresponding to cosmic void environments are represented with empty symbols while results for the entire simulation box are shown with filled symbols.

We plot the RVPF, $\chi$, as a function of the scaling variable, $\NE$, in Fig.~\ref{fig:chi_vs_NXi_bothspaces_invoid+box}. 
The right-side panels show the RVPFs in redshift space calculated for the entire box (solid red circles) and inside voids (empty red circles).
We find that there are no differences between the two environments, despite the low densities in voids by their definition.
Both curves closely follow the NB model of hierarchical clustering (solid grey line). 
The panels on the left, however, show that the RVPF in real space is indeed different inside voids (empty blue circles) as opposed to elsewhere in the box.
Throughout the box in real space (solid blue circles) the RVPF appears to somewhat follow the thermodynamical model (dash-dot grey line) but the fit is not thorough as it clearly overestimates the data for values of $\NE\gtrsim30$. 
On the other hand, the RVPF inside voids in real space follow the NB model surprisingly well.

We also plot the scales corresponding to the first and last values of the RVPF of each curve. These scales are indicated with vertical lines as a reference for the reader, since the corresponding scale for each value of $\NE$, i.e. the relation $\NE(R)$, will vary from sample to sample.
The bottom panels show the quantity $\Delta\chi_\mathrm{NB}\equiv\chi-\chi_\mathrm{NB},$ corresponding to the difference between the RVPF of the data and NB model for clarity.

\subsection{Random dilutions test}
Following the analysis of \cite{Croton2004models} we test the robustness of this result by randomly diluting the sample to verify whether the scaling relations found in Fig.~\ref{fig:chi_vs_NXi_bothspaces_invoid+box} hold. 
Results with random dilutions are shown in Fig.~\ref{fig:chi_vs_NXi_dilute_bothspaces_invoid+box}.
In both spaces (left and right panels) the top panels display the full sample throughout the simulation in coloured filled circles, and dilutions of 50, 25, and 10 per cent with squares, triangles, and crosses with incremental transparency respectively. 
The quantity $\Delta\chi_\mathrm{T}$ shown in the top left panel is defined in an analogous way as $\Delta\chi_\mathrm{NB}$ described above.
It can be seen that the dispersion between dilutions is larger in real space than in redshift space. This means that the scaling coefficients $S_p$ are not completely independent of the sample density, and therefore, hierarchical scaling in real space does not hold as rigorously as it does in redshift space.

We calculated the root-mean-square deviations of the data from the closest fit (the NB or thermodynamical model when computing $\Delta\chi_\mathrm{NB}$ or $\Delta\chi_\mathrm{T}$, respectively). For this calculation we only use data corresponding to $\NE\geq0.1$, given that, for smaller values, every model yields approximately the same $\chi(\NE)$, which is close to one. Notably, the rms deviations of the RVPF of galaxies in the general redshift space is one order of magnitude lower than that of real space galaxies with respect to their closest models.

It should be noted that larger dilutions shifts the range of $\NE$ to lower values, where the different models are more similar to each other and, consequently, rms values will tend to be lower. This means that larger dilutions probe a less statistically significant range of $\NE$ in the sense that it becomes more difficult to differenciate between models. Therefore, rms values should be regarded as a reference and compared amongst samples with the same dilution between both spaces. This is true also for the different analyses that follow in the next subsections.

We show the RVPF inside voids represented in the lower panels of Fig.~\ref{fig:chi_vs_NXi_dilute_bothspaces_invoid+box} with empty symbols. 
We can see that both spaces exhibit lower scatter within voids than elsewhere in the box, with similar rms values for each dilution in real and redshift space. 
This robustness with respect to dilutions reinforces the evidence in favour of hierarchical scaling within voids in both spaces.
This is unexpected given that previous works indicate that any hierarchical scaling found in redshift space breaks down in real space.

\begin{figure*}
    \centering
    \includegraphics[width=.8\textwidth]{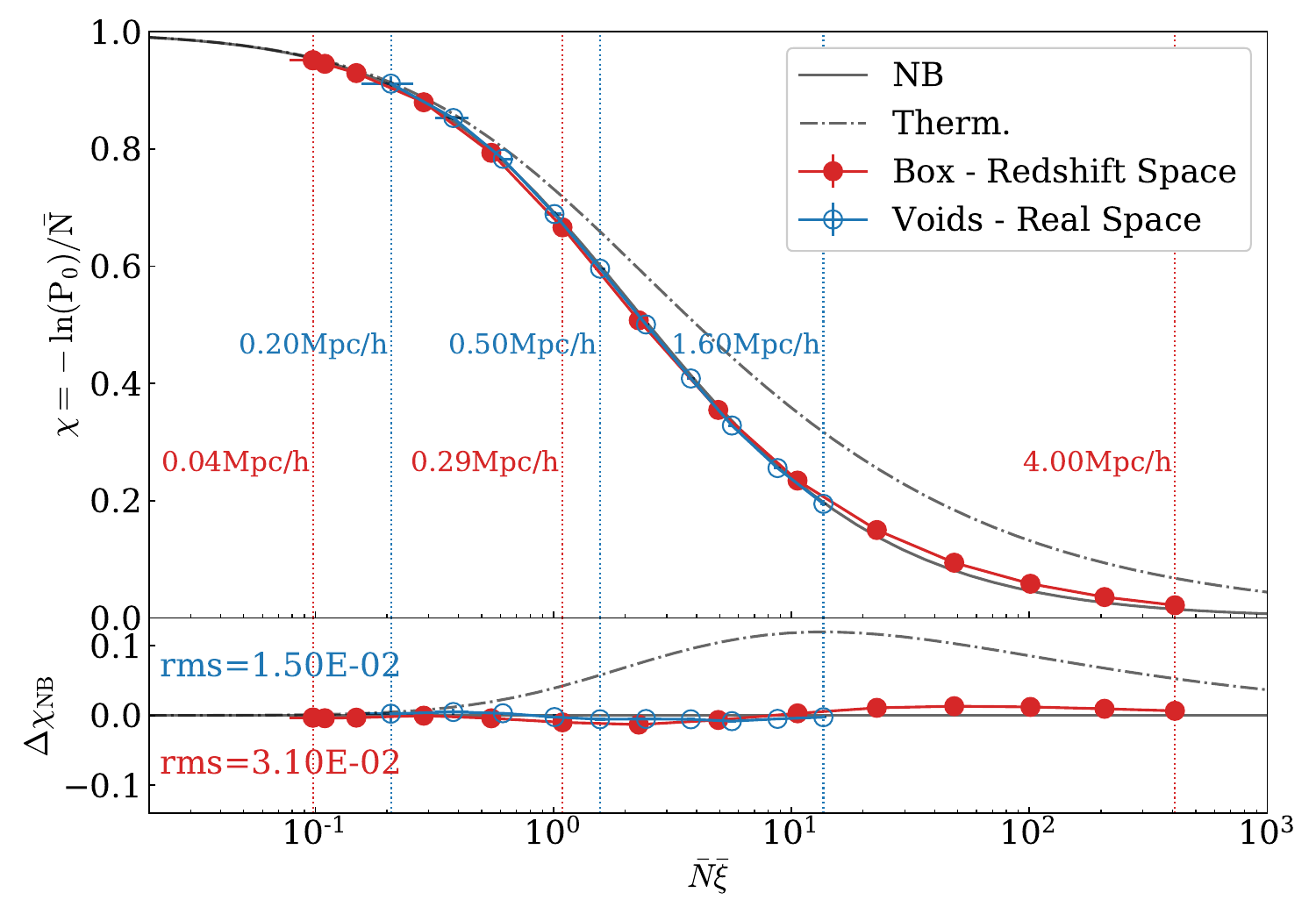}
    \caption{Reduced VPF, $\chi(\NE)$, inside voids (empty circles) and elsewhere in the simulation (filled circles), in real and redshift space (blue and red lines respectively). The negative binomial model of clustering (NB, black solid line) is known to be a good fit for the VPF of galaxies in redshift space. However, we find a striking agreement between this model, and the VPF of galaxies within voids in real space (empty blue circles), which is an indication of a previously unknown hierarchical scaling relation in real space.}
    \label{fig:chi_vs_NXi_rspaceinvoid_zspacebox}
\end{figure*}

Thus far we have found that the RVPFs for galaxies inside and outside voids are markedly different in real space, with the NB model being a very good fit for the RVPF inside voids.
On the other hand, we find no such difference in redshift space: both inside and outside voids, the RVPF is well fitted by the NB model. 
There is a striking agreement between the RVPFs of galaxies in the general redshift space and galaxies within real space voids (shown in Fig.~\ref{fig:chi_vs_NXi_rspaceinvoid_zspacebox}). This result indicates that within cosmic voids there is evidence of hierarchical scaling not found elsewhere in the Universe.

\subsection{VPF dependence on void identification parameters}
\label{sec:results_voidid}

From this section onwards, for clarity and visibility, we only show the zoomed-in differences between the calculated RVPFs and the NB or thermodynamic model accordingly ($\Delta\chi_\mathrm{NB}$ and $\Delta\chi_\mathrm{T}$), as shown in the lower panels of Figs.~\ref{fig:chi_vs_NXi_bothspaces_invoid+box} to~\ref{fig:chi_vs_NXi_rspaceinvoid_zspacebox}.

In this subsection we show how the RVPF changes with variations in the physical parameters that define a void: the minimum threshold of mass of the density tracers, $M_t$, and the integrated density contrast at one void radius, $\Delta(R_{\rm v})$. This allows us to gain insight into the effects on the RVPF of gradual change in the environment that is defined by a cosmic void. 

\subsubsection{Changing minimum mass of density tracers in void identification}

\begin{figure*}
    \centering
    \includegraphics[width=\textwidth]{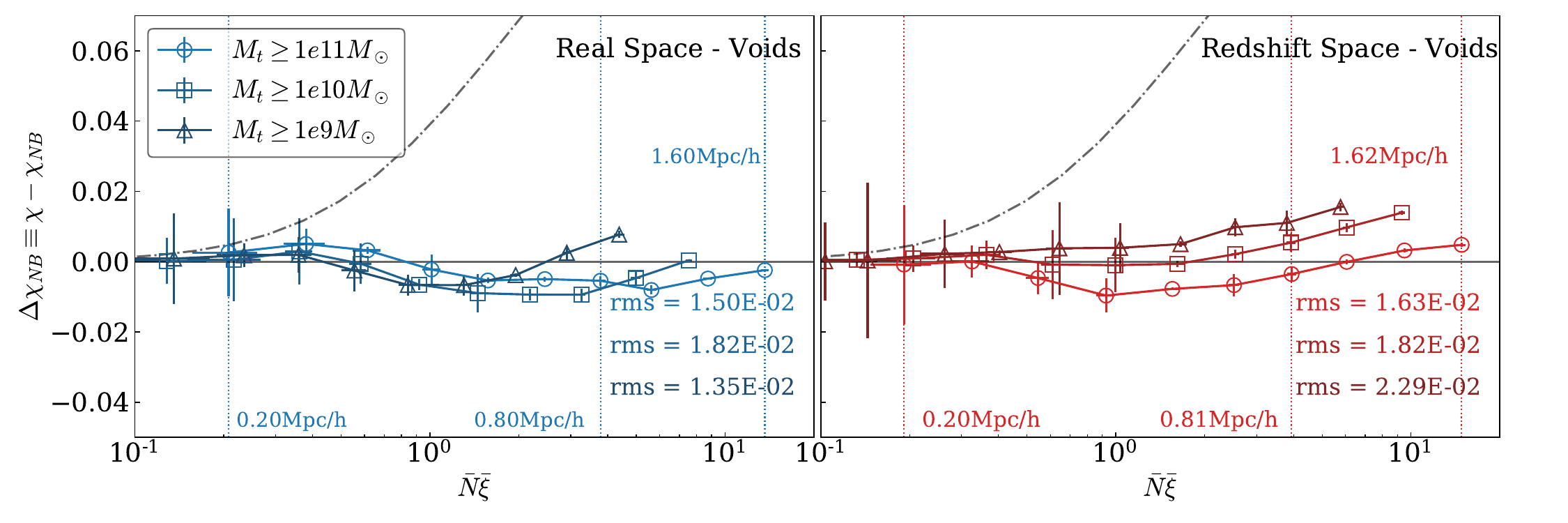}
    \caption{Reduced VPFs, plotted as differences to the NB model (solid grey line), of voids identified with different void tracer mass, $M_t$. Rms values are calculated for each sample w.r.t. the NB model and shown with the shade of colour of the corresponding curve (light to dark for larger to smaller $M_t$, respectively). We find no significant changes in the reduced VPFs, in either space, around the NB model. }
    \label{fig:rVPF_voidTracerMass}
\end{figure*}

We test the sensitivity of the RVPF to $M_t$, the mass of the density tracers used in void identification.
This is done by calculating the RVPF inside voids identified with different tracers (Fig.~\ref{fig:rVPF_voidTracerMass}). 
The main void sample (open circles) was identified using galaxies with masses greater than $10^{11}M_\odot$. The other lower mass thresholds we chose are $10^{10}M_\odot$ (open squares), $10^{9}M_\odot$ (open triangles).
The cutoff radius is 9, 7, and $6h^{-1}$Mpc for each void sample respectively.  
These choices in minimum void radii prevents including spurious or shot-noise voids in our void sample \citep{Correa2021}.
We find only small differences in the RVPFs when calculating them in voids traced by galaxies with different masses, with small variations of rms values with respect to the NB model in both spaces (rms $\simeq1$).
The RVPF and its agreement with the NB model seems to be insensitive to void tracer mass.

\subsubsection{Changing integrated density contrast}
\label{sec:results_intDelta}

\begin{figure*}
    \centering
    \includegraphics[width=\textwidth]{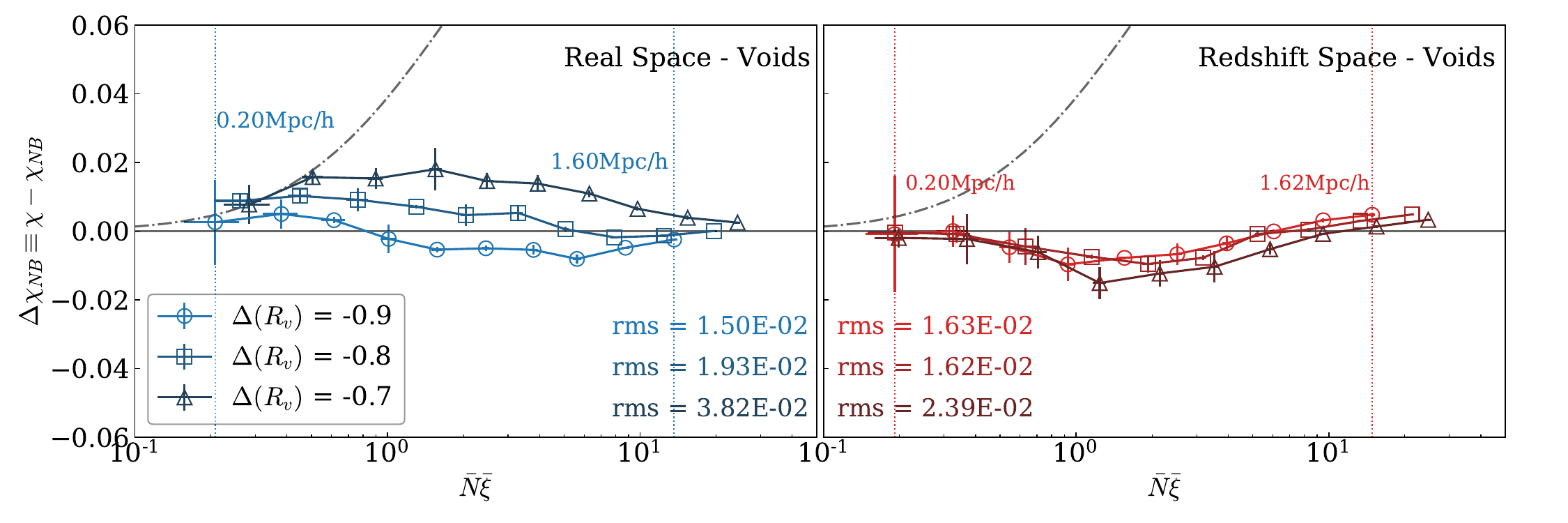}
    \caption{Reduced VPF plotted as differences to the NB model (solid grey line) in real and redshift space within voids identified with different integrated density contrast $\Delta(R_v)$.
    Rms values are calculated for each sample w.r.t. the NB model and shown with the shade of colour of the corresponding curve (light to dark for larger to smaller $\Delta(R_v)$, respectively).
    In real space we find a trend for the reduced VPF of voids with larger $\Delta(R_v)$ to be more like the reduced VPF throughout the box. This effect is not observed in redshift space, where the curves are statistically identical.
    }
    \label{fig:rVPF_intDelta}
\end{figure*}

Another important parameter in void identification is the integrated density contrast $\Delta(R_{\rm v})$.
It is standard practice to define a void as a sphere with $\Delta(R_{\rm v})=-0.9$ (see Sec.~\ref{sec:data}) so in addition we tested values of $\Delta(R_{\rm v})$= -0.8 and -0.7. Results for the RVPF in voids with each of these integrated deltas are shown in Fig.~\ref{fig:rVPF_intDelta}.

In real space, the structure as traced by voids with a lower density contrast, naturally yields higher values of RVPF, more similar to those found in the general Universe.
There is no such effect found in redshift space: the RVPF is robust against these integrated density changes used in the definition of voids. These results are consistent with Fig.~\ref{fig:chi_vs_NXi_dilute_bothspaces_invoid+box}.

\subsection{VPF dependence on mass}
\label{sec:results_mass}

\begin{figure*}
    \centering
    \includegraphics[width=\textwidth]{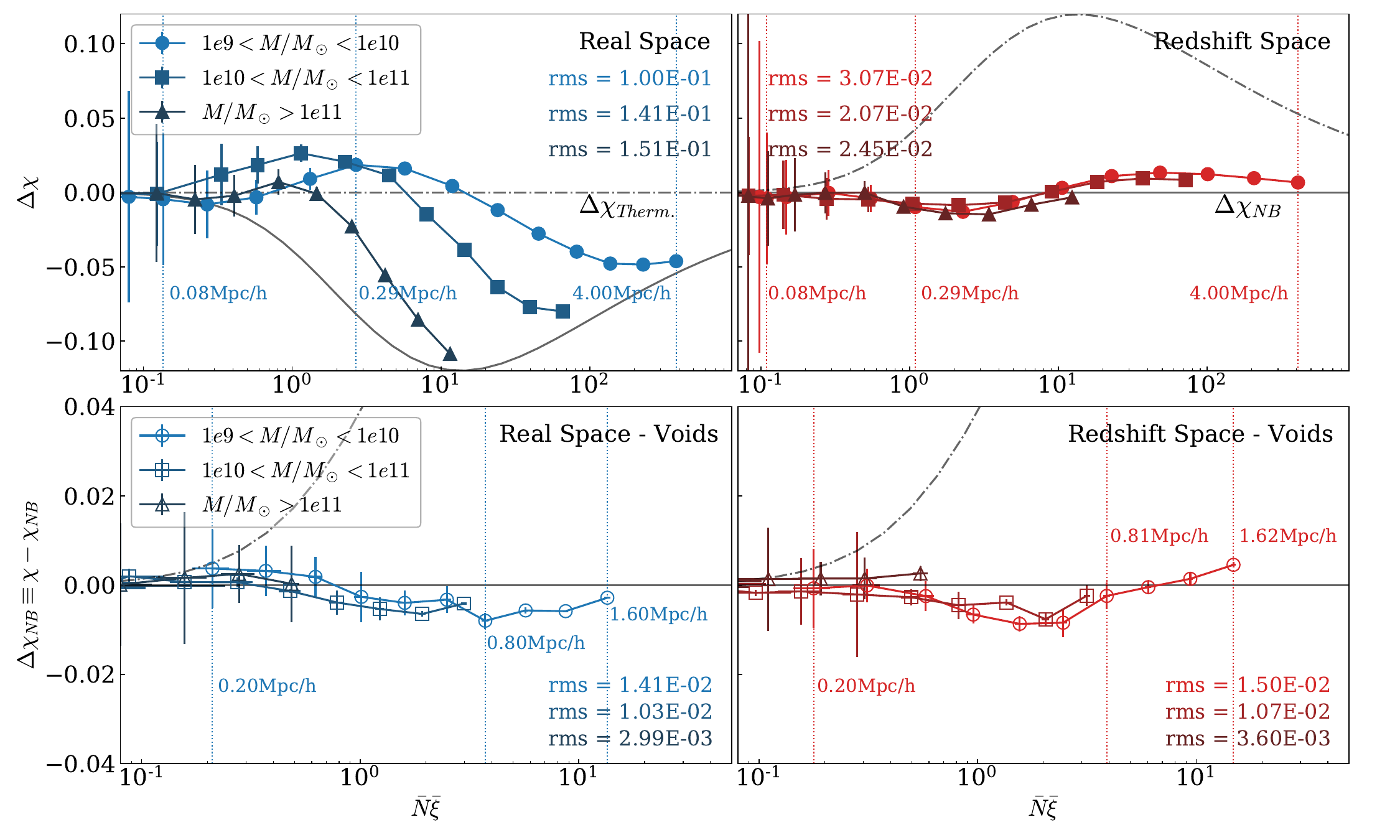}
    \caption{Reduced VPF plotted as the difference w.r.t. the NB model (solid grey line) calculated with different galaxy subsamples selected by mass ranges. 
    Rms values are calculated for each sample w.r.t. the NB model (solid grey line) and shown with the shade of colour of the corresponding curve (light to dark for smaller to larger mass values, respectively).
    Throughout the box (top panels) the differences between real and redshift space in selecting mass is stark: the hierarchical scaling detected in redshift space completely breaks down in real space.
    However, within voids (bottom panels), we detect hierarchical scaling in both spaces roughly consistent with the NB model.
    }
    \label{fig:rVPF_gxsMass}
\end{figure*}

Next, in Fig.~\ref{fig:rVPF_gxsMass}, we study the sensitivity of the VPF to galaxy mass.
We create galaxy subsamples according to the mass ranges $10^9$--$10^{10}M_\odot$ (circles), $10^{10}$--$10^{11}M_\odot$ (squares) and $M\geq10^{11}M_\odot$ (triangles). Then we calculate the RVPF throughout the box and inside the main void sample.

There is a clear difference in the effects of selecting galaxy mass between real and redshift space. While different subsamples fall on the same curve in redshift space, they have a wide spread in real space, further affirming the notion that hierarchical scaling is not found in the general Universe in real space. The scaling coefficients $S_p$ appear to have a strong dependence on mass in real space.

Within voids, however, the scenario is similar both in real and redshift space. Cosmic voids seem to provide a scenario where the $S_p$ appear to be independent of mass.

\subsection{VPF dependence on redshift}
\label{sec:results_z}

\begin{figure*}
    \centering
    \includegraphics[width=\textwidth]{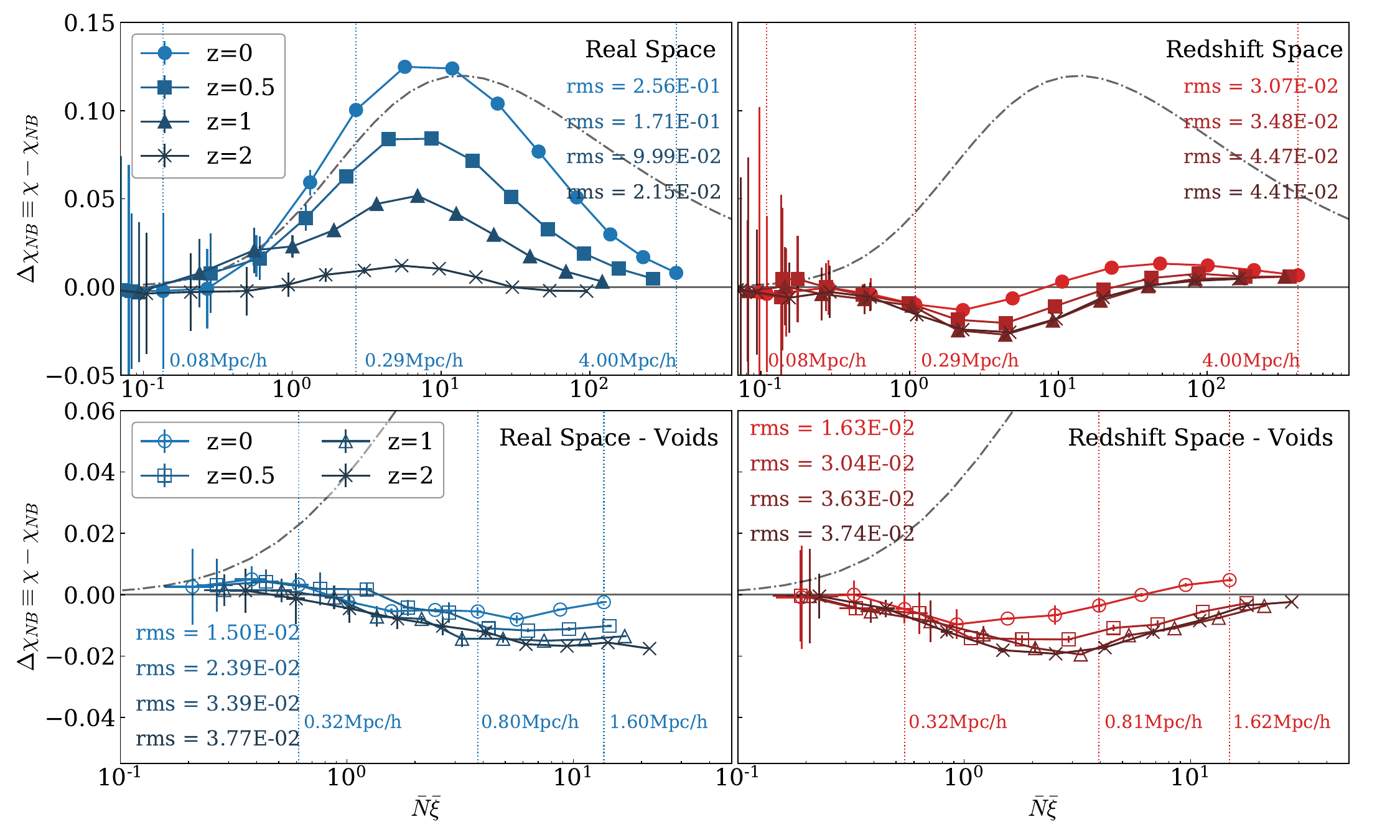}
    \caption{Reduced VPF plotted as differences to the NB model (solid grey line) calculated for the simulation different redshifts inside voids identified in different redshifts with their corresponding integrated density contrast.
    Rms values are calculated for each sample w.r.t. the NB model and shown with the shade of colour of the corresponding curve (light to dark for smaller to larger redshift values, respectively).
    Analogous to Fig.~\ref{fig:rVPF_gxsMass} we detect a great difference throughout the simulation (top panels) between real and redshift space for different redshifts where the RVPF in real space is much more affected by the evolution of structure than in redshift space.
    Within voids (bottom panels), the variation in the RVPFs is much smaller in both spaces, staying close the NB model in each case.
    }
    \label{fig:rVPF_redshift}
\end{figure*}

Finally, we want so study how the RVPF evolves with redshift.
We show the RVPF in real and redshift space at different redshift values, namely $z=0,0.5,1$ and 2, throughout the box and inside voids (Fig.~\ref{fig:rVPF_redshift}).
Following the spherical collapse model \citep{GunnGott1972,LiljeLahav1991,ShethvandeWeygaert2004}, for the different redshifts we identified voids with a time-dependent integrated density contrast value which corresponds to a perturbation with \mbox{$\Delta(R_{\rm v})=-0.9$} at $z=0$. For the mentioned redsfhit values, these are: $\Delta(R_{\rm v})\simeq-0.87$, -0.85, -0.80, respectively.

Throughout the box (top panels, Fig.~\ref{fig:rVPF_redshift}) we find a similar effect as with mass (Sec.~\ref{sec:results_mass}, Fig.~\ref{fig:rVPF_gxsMass}): the changes on the RVPF seen in redshift space are minor and all the curves are roughly fitted by the NB model, while in real space the differences are quite large. Interestingly, we detect a trend in real space for the RVPFs to be more similar to the NB model with increasing redshift. Between $z=0$ and $z=2$ (light blue circles and dark blue crosses, respectively) we have a difference of one order of magnitude in rms w.r.t the NB model. The RVPF of galaxies at high redshift in real space is similar to the RVPF of galaxies within voids at present time.

The computation of the RVPF inside voids at different redshifts (bottom panels) is more self-similar in both spaces, which would indicate that the $S_p$ have remained approximately constant with time. Naturally, the best fit within voids is still the NB model.

\section{Summary \& Discussion}
\label{sec:conclusions}

The VPF can be a pathway to higher-order correlations,
but a number of conditions need to be met. 
When the solutions of ``self-similarity'' and ``stable clustering'' are valid, then the scaling coefficients $S_p$ of Eq.~\ref{eq:hierarchicalscaling} are constant, and therefore the hierarchical scaling Ansatz holds and the reduced VPF, $\chi$, can be expressed in terms of the scaling variable, $\NE$, alone (Eq.~\ref{eq:rvpf_NE}). As a consequence, different samples fall on the same curve as long as they follow the same hierarchical clustering model.
The constancy of $S_p$ can be tested with random dilutions or by selecting subsamples with a given parameter of the sample and testing whether the RVPFs collapse onto the same curve. If they do not, it could be evidence that the $S_p$ have a strong dependence on said parameter, and therefore hierarchical clustering of the high-order correlations cannot be confirmed.

We have calculated the VPF in the TNG300-1 simulation, inside and outside of cosmic voids, both in real and redshift space. We compared the RVPF of galaxies in these samples with two popular clustering models: the negative binomial (NB) and the thermodynamic model. We have focused on the effects of the void environment on the VPF, and the sensitivity of this statistic with void definition, random sample dilution, galaxy mass, and structure evolution via redshift.
Previous works have shown that galaxies in redshift surveys are well fitted by the NB model of hierarchical clustering \mbox{\citep[e.g.][]{Croton2004models,Conroy2005}}, but that this fit breaks down in real space \citep{Lahav1993,Vogeley1994}.

We find that in redshift space, galaxies indeed appear to follow the NB model regardless of the environment. 
However, we report a strong difference in the RVPFs inside and outside cosmic voids in real space. 

Throughout the simulation, galaxies seem to follow the thermodynamic model up to $\NE\simeq30$, with an overall rms deviation of 0.10, although this fit does not hold when the sample is diluted. On the other hand, inside cosmic voids, we find an excellent agreement between the RVPF of galaxies and the NB model, with a rms of 0.01. This good fit is maintained even with dilutions.
For comparison, we find similar rms values (0.03 and 0.02) for the RVPFs of galaxies in redshift space throughout the box and inside voids respectively.

From this results it follows that the RVPF of galaxies inside cosmic voids in real space is very similar to the RVPF of galaxies throughout the box in redshift space, with both of them being well fitted by the NB model. I.e., the Ansatz of hierarchical scaling of the high-orders of the correlation function that has been shown to be valid for galaxies in redshift space can also be found in real space for galaxies within voids.

%RVPF void id
Following this analysis we study how the RVPF changes when changing what defines a void environment, which means calculating the RVPF inside voids for different integrated density contrasts, $\Delta(R_{\rm v})$, and different tracer mass, $M_t$, used to trace the underlying density when identifying cosmic voids.
We find that the RVPF is not sensitive to changes in $M_t$ in neither redshift nor real space, with the NB model still being the best fit.
On the other hand, while the RVPF remains insensitive to $\Delta(R_{\rm v})$ in redshift space, we notice a trend in real space where the data deviates from the NB model with larger density contrasts (i.e. more dense voids), and moves towards the RVPF of the general Universe in real space. Since we find that the RVPF in redshift space is the same regardless of the environment, it makes sense that we would find no significant changes in the RVPF with different $\Delta(R_{\rm v})$.

%RVPF mass
Next, we analysed the sensitivity of the RVPF to galaxy mass. In accordance to \cite{Croton2004models} we find that subsamples with different mass follow the NB model in redshift space, hinting towards hierarchical scaling. In real space, this agreement between the subsamples breaks down. However, we notice that the subsamples still follow the NB model inside voids both in real and redshift. 

%RVPF redshift
Finally, we studied how the RVPF changes with redshift, where, again, we detected the largest effects on the RVPF throughout the box in real space, and only small changes inside voids and everywhere in redshift space. Interestingly, we find that the RVPF in real space moves towards the NB model with increasing redshift. This leads to the NB model being a close fit for galaxies in cosmic voids at present time and galaxies at high redshift. This points towards the idea that the way in which high-order clustering emerges from the two-point correlation function in high redshifts is preserved within voids today.

%Discussion

The most widely used argument for the validity of hierarchical scaling in redshift space is that small-scale distortions dampen the amplitude of the coefficients $S_p$ so that they appear to be constant with scale (a necessary condition for hierarchical scaling), even though they are not in real space (see Fig.~49 of \citealt{Bernardeau2002}). This argument still warrants further exploration as to why the coefficients are affected by distortions in such a way that allows for apparent evidence of hierarchical scaling to arise. 
It could be the case that redshift space distortions blur strong non-linear effects at small scales and any other clustering behaviour that would break the hierarchical scaling relation. Cosmic voids and higher redshifts (where the VPF is closely fitted by the NB model, see Fig.~\ref{fig:chi_vs_NXi_rspaceinvoid_zspacebox} and \ref{fig:rVPF_redshift}) could provide an environment where velocity dispersions are low and, therefore, high non-linearities are still not a main agent in the clustering process, allowing for hierarchical scaling to be preserved.

It is also interesting to contrast our results with the discussion of \cite{Croton2006a} in which they argue that the peculiar motion of galaxies should not be the main agent in generating hierarchical scaling consistent with the NB model, as the results from \mbox{\cite{Vogeley1994}} might suggest. Instead, they find that faint red galaxies, which would be more affected by peculiar motions since they are in denser environments, are driven further away from (instead of towards) the NB model. 

Regarding the random dilutions analyses, it should be noted that the cosmic voids, by definition, are the least affected structures by this test. While this could partly be the reason why the statistics are less affected by dilutions within the cosmic voids, it should not be the main cause of the good agreement with hierarchical scaling. 
The RVPF by definition is disaffected from the mean density in each sampling sphere, which in principle makes it a good statistic to use in different environments.
Particularly, the maximum dilution is 10 per cent of the total sample, and is therefore comparable to the internal density of a cosmic void (see identification criteria and algorithms in \citealt{Ruiz2015}); and yet, even with this dilution, the RVPF outside cosmic voids is far from being well fitted by the NB model that seems to characterise the distribution inside voids in real space, as well as everywhere in redshift space.

Additionally, the success of the NB model in fitting the data is not well understood. 
One one hand, it has been shown that it is not a physically possible model since it breaks the second law of thermodynamics despite being the best fit for the data \citep{SaslawFang1996,YangSaslaw2011}, although some authors find its use justified \citep{CarruthersDuongVan1983,ElizaldeGaztanaga1992,Betancort-Rijo2009}.
Furthermore, the derivation of this phenomenological model (or any of the others) does not explicitly take redshift space distortions into account, so it has been surprising that it is such a good fit for the VPF of galaxies in redshift surveys and not in real space. However, in this work we find that this good fit is not a particularity of the redshift space. We show that it can also be found in real space within cosmic voids.

The theoretical background of the CiC statistics yields that if two VPFs are different, it is due not necessarily to the difference in sample density or two-point clustering amplitude (they could actually be identical), but a consequence of how the intrinsic clustering of the samples gives rise to a different hierarchical emergence of the higher-order correlations. For example, we find the VPFs in redshift space to be the same inside or outside the voids, although the clustering inside the voids is weaker and the density by definition is lower. However, in real space, the voids and the general Universe give markedly different VPFs. Under this framework, this difference is an effect of the high-order correlations.
Moreover, our results can also be understood as a consequence of the smaller changes in density contrast within voids.
The integrated density profiles of our voids are not completely flat within one void radius because we are identifying them with galaxies of higher mass ($M_t\geq10^{11}M_\odot$) than we are ultimately considering in the calculations ($M\geq10^9M_\odot$). Nevertheless, the changes in the density contrasts are smaller inside than outside voids.
Therefore, the solutions of self-similarity and stable clustering may be appropriate approximations in these underdense regimes.

Ultimately, in cosmic voids, we find evidence of hierarchical scaling that unravels further away in denser environments. The similarities between the RVPF in real space at higher redshifts and inside voids in present day indicate that clustering can be found inside voids that more pristinely preserve the initial conditions of the Universe. 
Upcoming surveys, such as \textit{Euclid} \citep{euclid} and the \textit{Legacy Survey of Space and Time} \citep[LSST,][]{lsst} at Vera Rubin Observatory, will have the required resolution and volume for a suitable statistical analysis with voids at higher redshifts, to contrast these results with observations.

\section*{Acknowledgements}

This work was partially supported by the Consejo Nacional de Investigaciones Cient\'ificas y T\'ecnicas, Argentina (PIP 11220200102832CO), the Secretar\'ia de Ciencia y Tecnolog\'ia, Universidad Nacional de C\'ordoba, Argentina, and the Agencia Nacional de Promoci\'on de la Investigaci\'on, el Desarrollo Tecnol\'ogico y la Innovaci\'on, Ministerio de Ciencia, Tecnolog\'ia e Innovaci\'on, Argentina.
The \textsc{IllustrisTNG} project used in this work (TNG300-1) have been run on the HazelHen Cray XC40-system at the High Performance Computing Center Stuttgart as part of project GCS-ILLU of the Gauss centres for Supercomputing (GCS).
This research has made use of NASA’s Astrophysics Data System. Visualizations made use of \textsc{python} packages.

%%%%%%%%%%%%%%%%%%%%%%%%%%%%%%%%%%%%%%%%%%%%%%%%%%
\section*{Data Availability}

 The data underlying this article will be shared on reasonable request to the corresponding author.



\bibliographystyle{mnras}
\bibliography{referencias}

\begin{thebibliography}{}
\makeatletter
\relax
\def\mn@urlcharsother{\let\do\@makeother \do\$\do\&\do\#\do\^\do\_\do\%\do\~}
\def\mn@doi{\begingroup\mn@urlcharsother \@ifnextchar [ {\mn@doi@} {\mn@doi@[]}}
\def\mn@doi@[#1]#2{\def\@tempa{#1}\ifx\@tempa\@empty \href {http://dx.doi.org/#2} {doi:#2}\else \href {http://dx.doi.org/#2} {#1}\fi \endgroup}
\def\mn@eprint#1#2{\mn@eprint@#1:#2::\@nil}
\def\mn@eprint@arXiv#1{\href {http://arxiv.org/abs/#1} {{\tt arXiv:#1}}}
\def\mn@eprint@dblp#1{\href {http://dblp.uni-trier.de/rec/bibtex/#1.xml} {dblp:#1}}
\def\mn@eprint@#1:#2:#3:#4\@nil{\def\@tempa {#1}\def\@tempb {#2}\def\@tempc {#3}\ifx \@tempc \@empty \let \@tempc \@tempb \let \@tempb \@tempa \fi \ifx \@tempb \@empty \def\@tempb {arXiv}\fi \@ifundefined {mn@eprint@\@tempb}{\@tempb:\@tempc}{\expandafter \expandafter \csname mn@eprint@\@tempb\endcsname \expandafter{\@tempc}}}

\bibitem[\protect\citeauthoryear{Balian, Schaeffer, Balian  \& Schaeffer}{Balian et~al.}{1989}]{BalianSchaeffer1989}
Balian R.,  Schaeffer R.,  Balian R.,   Schaeffer R.,  1989, A{\&}A, 220, 1

\bibitem[\protect\citeauthoryear{Baugh et~al.,}{Baugh et~al.}{2004}]{Baugh2004a}
Baugh C.~M.,  et~al., 2004, \mn@doi [Monthly Notices of the Royal Astronomical Society] {10.1111/j.1365-2966.2004.07962.x}, 351, L44

\bibitem[\protect\citeauthoryear{Benson, Hoyle, Torres  \& Vogeley}{Benson et~al.}{2003}]{Benson2003a}
Benson A.~J.,  Hoyle F.,  Torres F.,   Vogeley M.~S.,  2003, \mn@doi [Monthly Notices of the Royal Astronomical Society] {10.1046/j.1365-8711.2003.06281.x}, 340, 160

\bibitem[\protect\citeauthoryear{Bernardeau, Colombi, Gazta{\~{n}}aga  \& Scoccimarro}{Bernardeau et~al.}{2002}]{Bernardeau2002}
Bernardeau F.,  Colombi S.,  Gazta{\~{n}}aga E.,   Scoccimarro R.,  2002, \mn@doi [Physics Report] {10.1016/S0370-1573(02)00135-7}, 367, 1

\bibitem[\protect\citeauthoryear{Betancort-Rijo, Patiri, Prada  \& Romano}{Betancort-Rijo et~al.}{2009}]{Betancort-Rijo2009}
Betancort-Rijo J.,  Patiri S.~G.,  Prada F.,   Romano A.~E.,  2009, \mn@doi [Monthly Notices of the Royal Astronomical Society] {10.1111/J.1365-2966.2009.15567.X}, 400, 1835

\bibitem[\protect\citeauthoryear{{Carruthers} \& {Duong-van}}{{Carruthers} \& {Duong-van}}{1983}]{CarruthersDuongVan1983}
{Carruthers} P.,  {Duong-van} M.,  1983, \mn@doi [Physics Letters B] {10.1016/0370-2693(83)91103-6}, \href {https://ui.adsabs.harvard.edu/abs/1983PhLB..131..116C} {131, 116}

\bibitem[\protect\citeauthoryear{Ceccarelli, Padilla, Valotto  \& Lambas}{Ceccarelli et~al.}{2006}]{Ceccarelli2006}
Ceccarelli L.,  Padilla N.~D.,  Valotto C.,   Lambas D.~G.,  2006, \mn@doi [Monthly Notices of the Royal Astronomical Society] {10.1111/J.1365-2966.2006.11129.X}, 373, 1440

\bibitem[\protect\citeauthoryear{Coles \& Jones}{Coles \& Jones}{1991}]{ColesJones1991}
Coles P.,  Jones B.,  1991, \mn@doi [Monthly Notices of the Royal Astronomical Society] {10.1093/mnras/248.1.1}, 248, 1

\bibitem[\protect\citeauthoryear{Colombi, Bouchet  \& Schaeffer}{Colombi et~al.}{1994}]{Colombi1994AModel}
Colombi S.,  Bouchet F.~R.,   Schaeffer R.,  1994, \mn@doi [The Astrophysical Journal Supplement Series] {10.1086/192125}, 96, 401

\bibitem[\protect\citeauthoryear{Conroy et~al.,}{Conroy et~al.}{2005}]{Conroy2005}
Conroy C.,  et~al., 2005, \mn@doi [The Astrophysical Journal] {10.1086/497682}, 635, 990

\bibitem[\protect\citeauthoryear{{Correa}, {Paz}, {S{\'a}nchez}, {Ruiz}, {Padilla}  \& {Angulo}}{{Correa} et~al.}{2021}]{Correa2021}
{Correa} C.~M.,  {Paz} D.~J.,  {S{\'a}nchez} A.~G.,  {Ruiz} A.~N.,  {Padilla} N.~D.,   {Angulo} R.~E.,  2021, \mn@doi [\mnras] {10.1093/mnras/staa3252}, \href {https://ui.adsabs.harvard.edu/abs/2021MNRAS.500..911C} {500, 911}

\bibitem[\protect\citeauthoryear{Croton et~al.,}{Croton et~al.}{2004a}]{Croton2004models}
Croton D.~J.,  et~al., 2004a, \mn@doi [Monthly Notices of the Royal Astronomical Society] {10.1111/j.1365-2966.2004.07968.x}, 352, 828

\bibitem[\protect\citeauthoryear{Croton et~al.,}{Croton et~al.}{2004b}]{Croton2004}
Croton D.~J.,  et~al., 2004b, \mn@doi [Mon. Not. R. Astron. Soc] {10.1111/j.1365-2966.2004.08017.x}, 352, 1232

\bibitem[\protect\citeauthoryear{Croton, Norberg, Gaztanaga  \& Baugh}{Croton et~al.}{2006}]{Croton2006a}
Croton D.~J.,  Norberg P.,  Gaztanaga E.,   Baugh C.~M.,  2006, \mn@doi [Monthly Notices of the Royal Astronomical Society] {10.1111/j.1365-2966.2007.12035.x}, 379, 1562

\bibitem[\protect\citeauthoryear{Elizalde \& Gaztanaga}{Elizalde \& Gaztanaga}{1992}]{ElizaldeGaztanaga1992}
Elizalde E.,  Gaztanaga E.,  1992, \mn@doi [Monthly Notices of the Royal Astronomical Society] {10.1093/mnras/254.2.247}, 254, 247

\bibitem[\protect\citeauthoryear{Fry}{Fry}{1986}]{Fry1986}
Fry J.~N.,  1986, \mn@doi [The Astrophysical Journal] {10.1086/184747}, 308, L71

\bibitem[\protect\citeauthoryear{Fry \& Colombi}{Fry \& Colombi}{2013}]{Fry2013}
Fry J.~N.,  Colombi S.,  2013, \mn@doi [Monthly Notices of the Royal Astronomical Society] {10.1093/mnras/stt745}, 433, 581

\bibitem[\protect\citeauthoryear{Gaztanaga \& Yokoyama}{Gaztanaga \& Yokoyama}{1993}]{Gaztanaga1993}
Gaztanaga E.,  Yokoyama J.,  1993, \mn@doi [The Astrophysical Journal] {10.1086/172216}, 403, 450

\bibitem[\protect\citeauthoryear{{Gunn} \& {Gott}}{{Gunn} \& {Gott}}{1972}]{GunnGott1972}
{Gunn} J.~E.,  {Gott} J.~Richard I.,  1972, \mn@doi [\apj] {10.1086/151605}, \href {https://ui.adsabs.harvard.edu/abs/1972ApJ...176....1G} {176, 1}

\bibitem[\protect\citeauthoryear{{Hamilton}}{{Hamilton}}{1985}]{Hamilton1985}
{Hamilton} D.,  1985, \mn@doi [\apj] {10.1086/163537}, \href {https://ui.adsabs.harvard.edu/abs/1985ApJ...297..371H} {297, 371}

\bibitem[\protect\citeauthoryear{Hurtado-Gil, Mart{\'{i}}nez, Arnalte-Mur, Pons-Border{\'{i}}a, Pareja-Flores  \& Paredes}{Hurtado-Gil et~al.}{2017}]{Hurtado-Gil2017}
Hurtado-Gil L.,  Mart{\'{i}}nez V.~J.,  Arnalte-Mur P.,  Pons-Border{\'{i}}a M.~J.,  Pareja-Flores C.,   Paredes S.,  2017, \mn@doi [Astronomy and Astrophysics] {10.1051/0004-6361/201629097}, 601, A40

\bibitem[\protect\citeauthoryear{{Ivezi{\'c}} et~al.,}{{Ivezi{\'c}} et~al.}{2019}]{lsst}
{Ivezi{\'c}} {\v{Z}}.,  et~al., 2019, \mn@doi [\apj] {10.3847/1538-4357/ab042c}, \href {https://ui.adsabs.harvard.edu/abs/2019ApJ...873..111I} {873, 111}

\bibitem[\protect\citeauthoryear{{Jaeger} \& {Nagel}}{{Jaeger} \& {Nagel}}{1992}]{JaegerNagel1992}
{Jaeger} H.~M.,  {Nagel} S.~R.,  1992, \mn@doi [Science] {10.1126/science.255.5051.1523}, \href {https://ui.adsabs.harvard.edu/abs/1992Sci...255.1523J} {255, 1523}

\bibitem[\protect\citeauthoryear{{Lahav}, {Itoh}, {Inagaki}  \& {Suto}}{{Lahav} et~al.}{1993}]{Lahav1993}
{Lahav} O.,  {Itoh} M.,  {Inagaki} S.,   {Suto} Y.,  1993, \mn@doi [\apj] {10.1086/172143}, \href {https://ui.adsabs.harvard.edu/abs/1993ApJ...402..387L} {402, 387}

\bibitem[\protect\citeauthoryear{{Laureijs} et~al.,}{{Laureijs} et~al.}{2011}]{euclid}
{Laureijs} R.,  et~al., 2011, \mn@doi [arXiv e-prints] {10.48550/arXiv.1110.3193}, \href {https://ui.adsabs.harvard.edu/abs/2011arXiv1110.3193L} {p. arXiv:1110.3193}

\bibitem[\protect\citeauthoryear{{Lilje} \& {Lahav}}{{Lilje} \& {Lahav}}{1991}]{LiljeLahav1991}
{Lilje} P.~B.,  {Lahav} O.,  1991, \mn@doi [\apj] {10.1086/170094}, \href {https://ui.adsabs.harvard.edu/abs/1991ApJ...374...29L} {374, 29}

\bibitem[\protect\citeauthoryear{Marinacci et~al.,}{Marinacci et~al.}{2018}]{Marinacci2018}
Marinacci F.,  et~al., 2018, \mn@doi [Monthly Notices of the Royal Astronomical Society, Volume 480, Issue 4, p.5113-5139] {10.1093/MNRAS/STY2206}, 480, 5113

\bibitem[\protect\citeauthoryear{Maurogordato \& Lachieze-Rey}{Maurogordato \& Lachieze-Rey}{1987}]{Maurogordato1987VoidSegregation}
Maurogordato S.,  Lachieze-Rey M.,  1987, \mn@doi [Astrophysical Journal v.320, p.13] {10.1086/165520}, 320, 13

\bibitem[\protect\citeauthoryear{Mekjian}{Mekjian}{2007}]{Mekjian2007}
Mekjian A.~Z.,  2007, \mn@doi [The Astrophysical Journal] {10.1086/508151}, 655, 1

\bibitem[\protect\citeauthoryear{Naiman et~al.,}{Naiman et~al.}{2018}]{Naiman2018}
Naiman J.~P.,  et~al., 2018, \mn@doi [Monthly Notices of the Royal Astronomical Society, Volume 477, Issue 1, p.1206-1224] {10.1093/MNRAS/STY618}, 477, 1206

\bibitem[\protect\citeauthoryear{Nelson et~al.,}{Nelson et~al.}{2018}]{Nelson2018}
Nelson D.,  et~al., 2018, \mn@doi [Monthly Notices of the Royal Astronomical Society, Volume 475, Issue 1, p.624-647] {10.1093/MNRAS/STX3040}, 475, 624

\bibitem[\protect\citeauthoryear{Nelson et~al.,}{Nelson et~al.}{2019}]{Nelson2019b}
Nelson D.,  et~al., 2019, \mn@doi [ComAC] {10.1186/S40668-019-0028-X}, 6, 2

\bibitem[\protect\citeauthoryear{Padilla, Ceccarelli  \& Lambas}{Padilla et~al.}{2005}]{Padilla2005}
Padilla N.~D.,  Ceccarelli L.,   Lambas D.~G.,  2005, \mn@doi [Monthly Notices of the Royal Astronomical Society] {10.1111/J.1365-2966.2005.09500.X}, 363, 977

\bibitem[\protect\citeauthoryear{Peebles}{Peebles}{1980}]{Peebles1980}
Peebles P. J. E. P. J.~E.,  1980, {The large-scale structure of the universe}.
Princeton University Press

\bibitem[\protect\citeauthoryear{Pillepich et~al.,}{Pillepich et~al.}{2018}]{Pillepich2018b}
Pillepich A.,  et~al., 2018, \mn@doi [Monthly Notices of the Royal Astronomical Society, Volume 473, Issue 3, p.4077-4106] {10.1093/MNRAS/STX2656}, 473, 4077

\bibitem[\protect\citeauthoryear{Pillepich et~al.,}{Pillepich et~al.}{2019}]{Pillepich2019}
Pillepich A.,  et~al., 2019, \mn@doi [Monthly Notices of the Royal Astronomical Society, Volume 490, Issue 3, p.3196-3233] {10.1093/MNRAS/STZ2338}, 490, 3196

\bibitem[\protect\citeauthoryear{{Planck Collaboration} et~al.,}{{Planck Collaboration} et~al.}{2016}]{Collaboration2016}
{Planck Collaboration} et~al., 2016, \mn@doi [\aap] {10.1051/0004-6361/201525830}, \href {https://ui.adsabs.harvard.edu/abs/2016A&A...594A..13P} {594, A13}

\bibitem[\protect\citeauthoryear{{Ruiz}, {Paz}, {Lares}, {Luparello}, {Ceccarelli}  \& {Lambas}}{{Ruiz} et~al.}{2015}]{Ruiz2015}
{Ruiz} A.~N.,  {Paz} D.~J.,  {Lares} M.,  {Luparello} H.~E.,  {Ceccarelli} L.,   {Lambas} D.~G.,  2015, \mn@doi [\mnras] {10.1093/mnras/stv019}, \href {https://ui.adsabs.harvard.edu/abs/2015MNRAS.448.1471R} {448, 1471}

\bibitem[\protect\citeauthoryear{{Saslaw} \& {Fang}}{{Saslaw} \& {Fang}}{1996}]{SaslawFang1996}
{Saslaw} W.~C.,  {Fang} F.,  1996, \mn@doi [\apj] {10.1086/176949}, \href {https://ui.adsabs.harvard.edu/abs/1996ApJ...460...16S} {460, 16}

\bibitem[\protect\citeauthoryear{{Sharp}}{{Sharp}}{1981}]{Sharp1981}
{Sharp} N.~A.,  1981, \mn@doi [\mnras] {10.1093/mnras/195.4.857}, \href {https://ui.adsabs.harvard.edu/abs/1981MNRAS.195..857S} {195, 857}

\bibitem[\protect\citeauthoryear{{Sheth} \& {van de Weygaert}}{{Sheth} \& {van de Weygaert}}{2004}]{ShethvandeWeygaert2004}
{Sheth} R.~K.,  {van de Weygaert} R.,  2004, \mn@doi [\mnras] {10.1111/j.1365-2966.2004.07661.x}, \href {https://ui.adsabs.harvard.edu/abs/2004MNRAS.350..517S} {350, 517}

\bibitem[\protect\citeauthoryear{Springel}{Springel}{2010}]{Springel2010}
Springel V.,  2010, \mn@doi [Monthly Notices of the Royal Astronomical Society, Volume 401, Issue 2, pp. 791-851.] {10.1111/J.1365-2966.2009.15715.X}, 401, 791

\bibitem[\protect\citeauthoryear{Springel, White, Tormen  \& Kauffmann}{Springel et~al.}{2001}]{Springel2001}
Springel V.,  White S. D.~M.,  Tormen G.,   Kauffmann G.,  2001, \mn@doi [Monthly Notices of the Royal Astronomical Society, Volume 328, Issue 3, pp. 726-750.] {10.1046/J.1365-8711.2001.04912.X}, 328, 726

\bibitem[\protect\citeauthoryear{Springel et~al.,}{Springel et~al.}{2018}]{Springel2018}
Springel V.,  et~al., 2018, \mn@doi [Monthly Notices of the Royal Astronomical Society, Volume 475, Issue 1, p.676-698] {10.1093/MNRAS/STX3304}, 475, 676

\bibitem[\protect\citeauthoryear{Szapudi}{Szapudi}{1998}]{Szapudi1998}
Szapudi I.,  1998, \mn@doi [The Astrophysical Journal] {10.1086/305439}, 497, 16

\bibitem[\protect\citeauthoryear{Tinker, Weinberg  \& Warren}{Tinker et~al.}{2006}]{Tinker2006a}
Tinker J.~L.,  Weinberg D.~H.,   Warren M.~S.,  2006, \mn@doi [The Astrophysical Journal] {10.1086/504795}, 647, 737

\bibitem[\protect\citeauthoryear{Tinker, Conroy, Norberg, Patiri, Weinberg  \& Warren}{Tinker et~al.}{2008}]{Tinker2008}
Tinker J.~L.,  Conroy C.,  Norberg P.,  Patiri S.~G.,  Weinberg D.~H.,   Warren M.~S.,  2008, \mn@doi [The Astrophysical Journal] {10.1086/589983}, 686, 53

\bibitem[\protect\citeauthoryear{Vogeley, Geller, Park  \& Huchra}{Vogeley et~al.}{1994}]{Vogeley1994}
Vogeley M.~S.,  Geller M.~J.,  Park C.,   Huchra J.~P.,  1994, \mn@doi [The Astronomical Journal] {10.1086/117110}, 108, 745

\bibitem[\protect\citeauthoryear{White}{White}{1979}]{White1979}
White S. D.~M.,  1979, \mn@doi [Monthly Notices of the Royal Astronomical Society] {10.1093/mnras/186.2.145}, 186, 145

\bibitem[\protect\citeauthoryear{{Yang} \& {Saslaw}}{{Yang} \& {Saslaw}}{2011}]{YangSaslaw2011}
{Yang} A.,  {Saslaw} W.~C.,  2011, \mn@doi [\apj] {10.1088/0004-637X/729/2/123}, \href {https://ui.adsabs.harvard.edu/abs/2011ApJ...729..123Y} {729, 123}

\makeatother
\end{thebibliography}

%%%%%%%%%%%%%%%%%%%%%%%%%%%%%%%%%%%%%%%%%%%%%%%%%%

%%%%%%%%%%%%%%%%% APPENDICES %%%%%%%%%%%%%%%%%%%%%

\appendix

\section{Consistency and Stability of the main CiC quantities}
\label{sec:appA}

\begin{figure}
    \centering
    \includegraphics[width=\columnwidth]{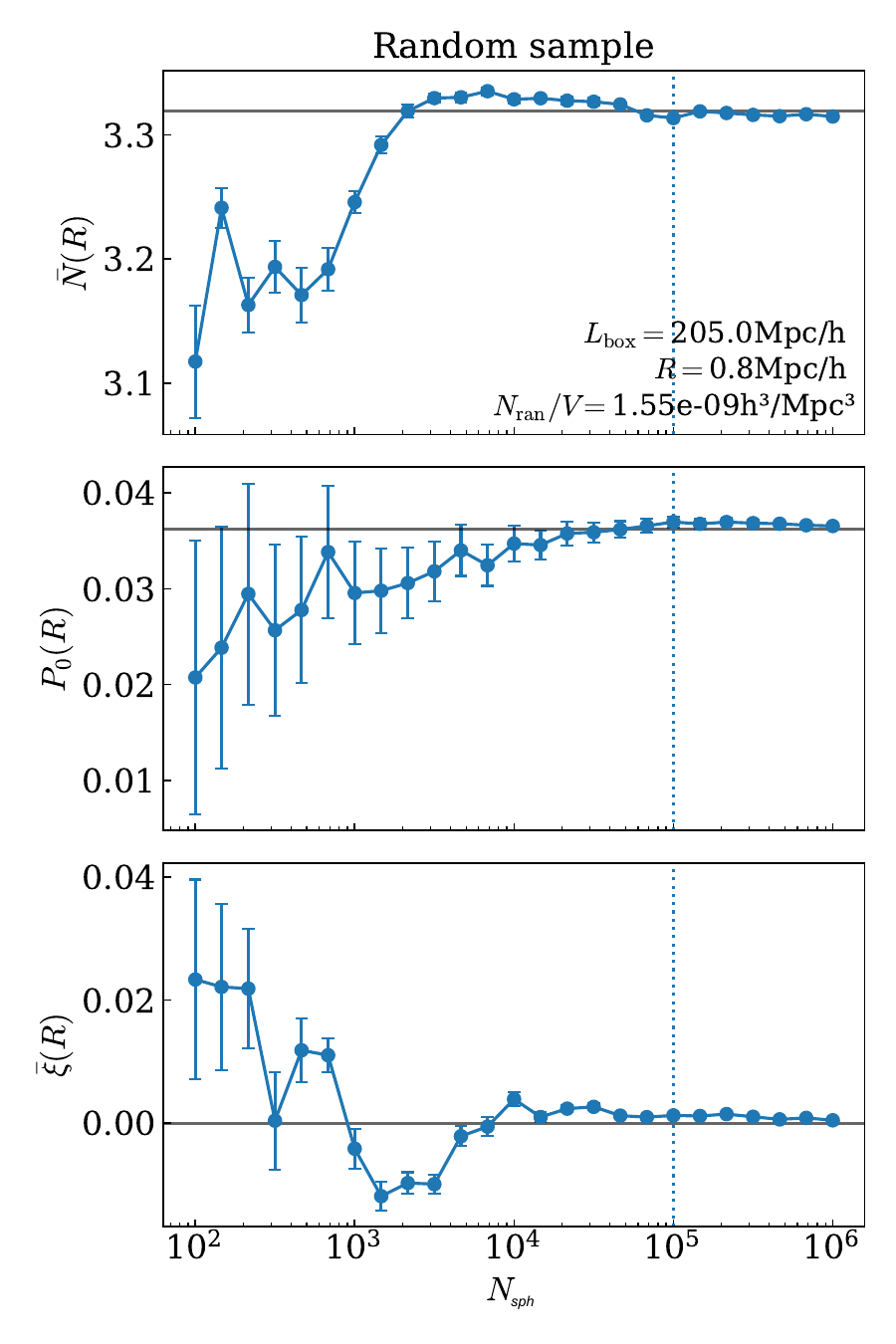}
    \caption{Consistency and stability test for the CiC statistics $P_0$, $\bar{N}$ y $\bar{\xi}$ as a function of the number of test spheres $N_\mathrm{sph}$. The horizontal solid lines indicate the theoretical value of the statistics for a random sample with the same number density as the TNG300-1 simulation after selecting galaxies with $10^9\geq M/M_{\sun}\geq10^{13}$. The vertical dotted line shows the value of $N_\mathrm{sph}$ chosen for this work as the number of test spheres for a reliable estimation of the statistics.}
    \label{fig:stabilitytests}
\end{figure}

Following the analysis of \citealt{Conroy2005}, we study the sensitivity of $P_0$, $\Bar{N}$, and $\Bar{\xi}$ (the main CiC statistics used in this work) to the number of test spheres $N_\mathrm{sph}$. This test will show if our algorithm for calculating the VPF indeed is correct, and will inform on how many test spheres is necessary for a reliable estimation of the aforementioned quantities.
We confirm that the estimation of the statistics stabilizes around the expected value, and see how many test spheres are necessary for this stabilization to ocurr.

We perform the calculations on a randomly distributed sample of points with the same density as the largest data used in this work: the TNG300-1 galaxy sample after selecting those with $10^9\geq M/M_{\sun}\geq10^{13}$. This sample size is approximately $10^7$, with a mean interparticle distance of $\simeq0.86h^{-1}$Mpc. Therefore we fix our testing radius slightly smaller at $R=0.8h^{-1}$Mpc so as to ensure a value of $P_0(R)\ne0$. Results are shown in Fig.~\ref{fig:stabilitytests} with uncertainties calculated as described in Sec.~\ref{sec:uncertainties}.

As expected, for small $N_\mathrm{sph}$ these quantities are unstable but eventually converge at the theoretical values marked with horizontal grey lines. 
To ensure a reliable estimate of these quantities, we use $N_\mathrm{sph}=10^5$ (vertical dotted blue lines) throughout this work, although we have obtained robust results within error bars with as few as $\simeq10^3$ test spheres.

%%%%%%%%%%%%%%%%%%%%%%%%%%%%%%%%%%%%%%%%%%%%%%%%%%

% Don't change these lines
\bsp	% typesetting comment
\label{lastpage}
\end{document}